\documentstyle [times,cite,epsf,twoside,11pt,here]{article}

 \def\PgSm{\ifmmode\mathrm{\Sigma^-}
           \else$\mathrm{\Sigma^-}$\fi}
 \def\PgL{\ifmmode\mathrm{\Lambda}
          \else$\mathrm{\Lambda}$\fi}
 \def\PgXm{\ifmmode\mathrm{\Xi^-}
           \else$\mathrm{\Xi^-}$\fi}
 \def\PgOm{\ifmmode\mathrm{\Omega^-}
           \else$\mathrm{\Omega^-}$\fi}
 \def\PK{\ifmmode\mathrm{K}
         \else$\mathrm{K}$\fi}
 \def\PKpm{\ifmmode\mathrm{K^{\pm}}
           \else$\mathrm{K^{\pm}}$\fi}
 \def\PKp{\ifmmode\mathrm{K^+}
          \else$\mathrm{K^+}$\fi}
 \def\PKm{\ifmmode\mathrm{K^-}
          \else$\mathrm{K^-}$\fi}
 \def\Pgp{\ifmmode\mathrm{\pi}
          \else$\mathrm{\pi }$\fi}
 \def\Pgpm{\ifmmode\mathrm{\pi^-}
           \else$\mathrm{\pi^-}$\fi}
 \def\Pgpp{\ifmmode\mathrm{\pi^+}
           \else$\mathrm{\pi^+}$\fi}
 \def\Pgppm{\ifmmode\mathrm{\pi^{\pm }}
            \else$\mathrm{\pi^{\pm }}$\fi}
 \def\Pgpz{\ifmmode\mathrm{\pi^0}
           \else$\mathrm{\pi^0 }$\fi}
 \def\Pe{\ifmmode\mathrm{e}
         \else$\mathrm{e}$\fi}
 \def\Pep{\ifmmode\mathrm{e^+}
          \else$\mathrm{e^+}$\fi}
 \def\Pem{\ifmmode\mathrm{e^-}
          \else$\mathrm{e^-}$\fi}
 \def\Pgm{\ifmmode\mathrm{\mu}
          \else$\mathrm{\mu}$\fi}
 \def\Pgmm{\ifmmode\mathrm{\mu^-}
           \else$\mathrm{\mu^-}$\fi}
 \def\Pgmp{\ifmmode\mathrm{\mu^+}
           \else$\mathrm{\mu^+}$\fi}

\def\kevc1{\ifmmode\mathrm{\ keV/{\mit c}}
          \else$\mathrm{\ keV/{\mit c}}$\fi}
\def\Mevc1{\ifmmode\mathrm{\ MeV/{\mit c}}
          \else$\mathrm{\ MeV/{\mit c}}$\fi}
\def\gevc1{\ifmmode\mathrm{ GeV/{\mit c}}
          \else$\mathrm{ GeV/{\mit c}}$\fi}
\def\kevc2{\ifmmode\mathrm{\ keV/{\mit c}^2}
          \else$\mathrm{\ keV/{\mit c}^2}$\fi}
\def\Mevc2{\ifmmode\mathrm{\ MeV/{\mit c}^2}
          \else$\mathrm{\ MeV/{\mit c}^2}$\fi}
\def\Gevc2{\ifmmode\mathrm{\ GeV/{\mit c}^2}
          \else$\mathrm{\ GeV/{\mit c}^2}$\fi}
\def\Gev2c2{\ifmmode\mathrm{\ GeV^2/{\mit c}^2}
          \else$\mathrm{\ GeV^2/{\mit c}^2}$\fi}

\def\xf{\ifmmode x_{\mathrm{F}}
          \else$ x_{\mathrm{F}}$ \fi}

\def\Pp{\ifmmode{\mathrm p}
         \else${\mathrm p}$\fi}
\def\Pn{\ifmmode{\mathrm n}
         \else${\mathrm n}$\fi}
\newcommand{\dcy}{\mbox{$ \rightarrow $}}

\begin{document}
\vspace*{0.8cm}
\begin{flushright}
\today
\end{flushright}
\vspace{0.8cm}
\Large
           
\centerline{The high-intensity hyperon beam at CERN}
\large
\vspace{1.8cm}
\vspace{0.4cm}
\normalsize

\noindent
\sloppy
  Yu.A.Alexandrov$^1$, 
  M.Clement$^2$, 
  F.Dropmann$^3$,
  A.Fournier$^4$
  P.Grafstr\"om.$^2$, 
  E.Hubbard$^3$,
  S.Paul$^{3,a}$, 
  H.W.Siebert$^5$, 
  A.Trombini$^3$ 
  M.Zavertiaev$^1$ 
\vspace{0.2cm}
\begin{flushleft}
\noindent
    $^1${\sl Moscow Lebedev Physics Inst., RU-117924, Moscow, Russia.}\\
    $^2${\sl CERN; CH-1211 Gen\`eve 23, Switzerland.}\\
    $^3${\sl Heidelberg Max-Planck-Inst. f\"ur Kernphysik$^\#$; 
          Postfach 103980, 
          D-69029 Heidelberg, Germany.}\\
    $^4${\sl Grenoble ISN; 53 Avenue des Martyrs ; F-38026 Grenoble
    CEDEX, France } \\
    $^5${\sl Heidelberg, Univ., Physikal. Inst.$^\#$; 
          Philosophenweg 12, 
         D-69120 Heidelberg, Germany.}\\
\end{flushleft}
\vspace{0.2cm}

\normalsize
\centerline{\underline{\large{Abstract}}}
\vspace{0.5cm}
A high-intensity hyperon beam was constructed at CERN to 
deliver \PgSm\ to experiment WA89 at the Omega facility and 
operated from 1989 to 1994. The setup allowed rapid changeover between
hyperon and conventional hadron beam configurations.  
 The beam provided  a \PgSm\ flux of $1.4 \times 10^5$ per burst 
at  mean momenta between 330 and 
345 \gevc1\,  produced by about  $3 \times 10^{10}$ protons of 450 \gevc1 .
At the experiment target the beam had a \PgSm/\Pgpm\ ratio 
close to 0.4 and a size of 1.6 $\times $ 3.7 cm$^2$.   
The beam particle trajectories and their momenta were measured 
with a scintillating fibre hodoscope in the beam channel
and a silicon microstrip detector at the exit of the channel. 
A fast transition radiation detector was used to identify the pion component
of the beam. 

\vspace{2cm}

\centerline{(Submitted to Nuclear Instruments and Methods A)}
\vspace{6.0cm}

\hrule width 6truein
\vspace{0.2cm}
\small
\noindent
$a$) Now at now at Technische Universit\"at M\"unchen,Graching,
  Germany \\
{$\#$) supported by the Bundesministerium f\"ur Bildung,
  Wissenschaft,Forschung und Technologie, 
  Germany, under contract numbers 05~5HD15I and  06~HD524I }

\setlength{\oddsidemargin}   {0.0in}
\setlength{\evensidemargin}  {0.0in}
\setlength{\textwidth}       {6.5in}
\setlength{\textheight}      {9.0in}
\setlength{\topmargin}      {-0.3in}
\setlength{\headheight}      {0.3in}
\setlength{\headsep}         {0.3in}
\setlength{\footskip}        {0.6in}
\setlength{\footheight}      {0.3in}

\normalsize

\clearpage
\noindent
\setcounter{page}{1}
\section{Introduction}
\pagestyle{plain}

 Our present experimental knowledge of the properties and 
interactions of hyperons
comes mainly from hyperon beam experiments. 
In contrast to normal hadron beam lines, hyperon beams are characterized 
by very short lengths of order 10m, which represent a compromise
between the need for adequate shielding  between the primary
hyperon production target and the experimental apparatus and 
the need to reduce decay losses of the hyperons
passing along the beam line.
At a momentum of 100 \gevc1, the hyperon decay lengths vary
from 7.1m for $\Lambda$ to 1.5m for $\Omega^-$,
which leads to severe reductions of the hyperon fluxes available for experiments.
This has induced a continuous quest for higher hyperon beam momenta,
which started at about 25 \gevc1 at the Brookhaven AGS and the CERN PS in the
early seventies.
The advent of the super proton synchrotrons at CERN and at FERMILAB in
the mid-seventies
led to the immediate construction of new hyperon beam lines
to exploit the newly available high energies. 

At CERN, a charged hyperon beam with momenta between 70 and 135 \gevc1
was operated between 1976 and 1982 in the CERN West Hall.
It achieved fluxes of 12,000 $\Sigma^-$, 400 $\Xi^-$ and 10 $\Omega^-$
per beam pulse, all of which were successfully used in experiments.
For reviews of these activities see \cite{bo84} and \cite{pon85}.

 The quest for higher $\Sigma^-$ fluxes then led to the construction of a new
hyperon beam line at CERN, capable of delivering in excess of 10$^5$
$\Sigma^-$  per beam pulse at 340 \gevc1  to the Omega spectrometer facility
in the West Hall. The beam was brought
into operation in 1989-90 and was used for physics runs in 1991, 1993 and 1994.
The design and operation of the beam are the subject of this article.

At FERMILAB  the development of hyperon beams 
culminated in a 600 \gevc1\ beam delivering \PgSm\ to the 
SELEX experiment in 1997 \cite{selex}.

\section{Design considerations}\label{sec:design}

\subsection{ Boundary conditions}

 Budget and manpower restrictions led to the decision to base the 
hyperon beam experiment WA89 on the existing Omega spectrometer 
facility in the CERN West Area \cite{Proposal}, which was used by
several different experiments each year.
This imposed quite severe restrictions on the 
design of the beam, which are listed below:

\begin{itemize}
\item The Omega spectrometer was situated at the end of the H1 beam line in the West 
  Area. H1 is a high energy hadron/electron beam used by a number of different 
  experiments. The new 
  hyperon beam had to be designed allowing   for a short changeover time 
  between the different  experiments and the different modes of
  operating H1. 
\item In addition to  the H1 beam there are five more beams in the West Area complex 
  operating  at the same time. A hyperon beam always implies dumping 
  a large number of protons close to the experiment, which creates an 
  unavoidable muon background in the experimental hall. It was mandatory to 
  reduce the muon flux to the neighbouring 
  experiments as much as possible.
\item The West Area, where the Omega spectrometer was located,
  is an open unshielded hall  . It was 
  necessary to cover more than 100 m of the H1 beam 
  and in particular the surroundings of the hyperon production target with shielding.
  The maximum amount of shielding that could be installed without interfering 
  too much with existing installations limited the proton flux to   
  $5\times10^{10}$ protons 
  per pulse thereby imposing a limit on the intensity of the hyperon beam.
\item Economically it was inconceivable to build any new beam line magnets and we 
  thus had to rely upon those  magnetic elements being available.  
\end{itemize}

\subsection{Basic design}\label{sec:basic}

The conditions listed above  left little freedom in the basic design \cite{Basic}.
The first condition made a wobbling  configuration  a natural choice in order to  
preserve the  direction of the H1 line and not to be forced to move the Omega 
spectrometer. The second condition suggested a  bending of the beam in the 
vertical plane in order not to sweep the muons horizontally into other beam lines. 
Condition three together with the wanted \PgSm\ flux fixed the acceptance of the 
hyperon channel. The fourth condition limited the bending power available for the beam,
the strongest bending magnets 
available being three H-type magnets, 3.6 meters long and each with a maximum 
field of 2.4 T.
 
At an early stage it was decided not to use any quadrupoles. This decision was 
dictated by space limitations and by the loss of intensity due to an increased 
beam length. 

The choice of the beam momentum was a compromise between conflicting
requirements.
On the one hand, the $\Sigma^- / \pi^-$ production
ratio increases rapidly with  the ratio of the hyperon beam momentum p$_Y$
and the primary proton momentum, $ \xf \approx $p$_Y$/p$_p$ with
p$_p = 450 $ \gevc1  , 
and the \PgSm\ decay losses are reduced by increasing p$_Y$.
On the other hand, the \PgSm\ production cross section
decreases rapidly with \xf\ as $\sigma \propto (1-\xf)^3$.
The limits on the primary proton flux mentioned above led us to the
choice of p$_c = 355$ \gevc1\ for the momentum on the central trajectory
of the collimator, which corresponds to \xf = 0.8.

Fig. \ref{fig:1} shows the basic design of the beam line. 
The 450 \gevc1 protons are incident from the left and bent 
down 7 mrad before interacting in the production target.  Downstream of the 
target there were  three 2.4 T magnets all filled with brass and tungsten 
collimators. A channel with small transverse dimensions was embedded in the 
tungsten defining the passage of the beam. The first magnet contained the dump 
region discussed below. The momentum spread of the 
beam was defined by the channel in the second magnet. The third magnet bent the 
beam back onto the geometrical axis of the Omega spectrometer.

\subsection{Flux calculations}\label{sec:predict}

The \PgSm\ flux of the beam was estimated using the known 
\Pgpm-flux and measured values 
of the \PgSm/\Pgpm ratio as a function of $\xf$ (see
Fig. \ref{fig:2}). For the \Pgpm\ flux we used 
measurements in the H1 and H8 beams both made at $\xf=0.8$. As the momentum 
bite of the beam is quite large the $\xf$ dependence of the \Pgpm\
flux also enters into the flux calculation. Here we assumed an
$\xf$-dependence of the form $(1-\xf)^4$. The 
result of the flux calculation as a function of the central momentum of the 
channel is shown in Fig. \ref{fig:3}. The basic design is compared with a solution without 
wobbling. Also shown in figure \ref{fig:3} is the gain 
that would have been achieved with 3.0 T magnets
had they been available. With these stronger magnets the required
momentum spread of   $\delta p/p =  15 \% $ (FWHM) would have 
allowed for a larger angular acceptance of the magnetic channel and therefore 
an increased flux. This increase in flux is also seen in fig. \ref{fig:3}. 
At the chosen value of the central momentum, p$_c = 355 $ \gevc1  ,
the calculated \PgSm/\Pgpm\ ratio was somewhat below 1.
The design parameters of the beam are given in table \ref{tab:1}.

\section{Layout}\label{sec:layout}

\subsection{Proton beam}{\label{sec:proton}

The primary proton beam of the SPS with an energy of 450 \gevc1\ was extracted  to 
the West Area in a slow extraction mode with a spill duration of  2.6 s and a 
cycle time of 14.4 s. The intensity was typically  a few times $10^{12}$ 
protons per spill. A first intensity attenuation  of the primary beam was provided  
by a production target (T1). This 500 mm long Be target  produced secondaries 
for the neighbouring H3 beam. A further reduction in intensity by roughly a factor 
10  was obtained by  a fixed hole collimator \mbox{20 m} downstream of the T1 target. 
This collimator consisted of 4 iron blocks with transverse dimensions
of  \mbox{ 80 cm $\times$ 120 cm}   and with a length of 40 cm each in the
direction of the beam. 
The blocks  were equipped with tungsten inserts having small "pinholes" of diameter 3.3 mm. 
The fine tuning of the intensity was achieved by  a movable four-jaw collimator 
located 35 m further downstream. It should be pointed out that  the T1 target and 
the subsequent collimators were located  in the horizontal plane of the accelerator 
ring i.e. some 20 m below the surface. This means that the muon background 
coming from the proton beam collimation is negligible at the surface, where the 
hyperon beam and the experiment were located.

The distance between the T1 target and the hyperon production target  was
approximately  800 meters. The protons were transported over this distance via the H1 
beam line. This was a general purpose, high energy and high resolution beam (for 
details see \cite{Westbeam} ). The last 100 meters of the beam were modified in order to make 
space for the hyperon beam upstream of the Omega spectrometer.

The H1 beam line  is  equipped with MWPCs and scintillator filament scanners 
normally used for beam tuning . Obviously, those detectors could not be used for 
beam intensities of several $10^{10}$ protons per pulse. In order to tune the 
proton beam, attenuators consisting of  120 cm Be, i.e three interaction lengths, 
could be introduced  just downstream of the T1 target. Together with the 
abovementioned four-jaw collimator  this allowed to reduce the intensity down to  
the range of $10^7$ protons per pulse  suitable for tuning .
Once the beam was tuned the absorbers were taken away and the collimators 
opened to obtain the full proton intensity on the hyperon production 
target. Several types of secondary emission monitors were installed just in front of 
the  target in order to measure and monitor the proton flux and the position of the 
proton beam on the target. Fig. \ref{fig:4}.a and \ref{fig:4}.b show the horizontal and vertical 
profiles of the proton beam as measured with one of the  secondary emission 
monitors  just upstream of the target. The horizontal and vertical  spot sizes were
$250 \mu$m and $450 \mu$m, respectively (standard deviation). The divergence of the 
proton beam at the target was about 0.1 mrad horizontally and vertically.

\subsection{Target choice}\label{sec:target}

The target consisted of four Be rods each 100 mm long and 2 mm in diameter. 
The rods were suspended on thin aluminium foils ($ 25 \mu m$)  at
each end. The target was motorized and could be moved in  two 
orthogonal directions in a plane transverse to the beam.
 There is an interest to keep the diameter of the target as small as possible as long 
as the proton beam is well inside the target diameter. This minimizes secondary 
interactions which create low energy pions subsequently decaying into muons. 
The proton  beam was well contained within the diameter of the target as can be 
seen from figures \ref{fig:4}.a and \ref{fig:4}.b .
Beryllium was chosen as target material because the total inelastic cross section 
increases faster with A than the leading particle cross section. This means that, per 
interacting proton, the fraction of hyperons is largest for low A targets.

\subsection{Magnetic channel}\label{sec:magnetic}

Fig. \ref{fig:1c} shows the shape of the magnetic  channel 
in both the bend and non-bend 
plane. Note that the horizontal axis is in meters and the vertical in cm. The 
channel is embedded in tungsten. Tungsten is chosen due to its very short 
interaction length thus stopping as  many particles as possible  before the decay.

The basic shape in both planes is a wedge opening towards the exit. Some 
tolerance has been added to take into account possible misalignments between 
successive magnets. 

The shape of the first part of the channel is designed trying to minimise the muon 
background that originates from secondary pions and kaons created in the 
production target and decaying into muons before being absorbed in the walls of the 
channel.  It turns out that the lateral dimensions are of small importance and what 
really matters is the longitudinal distance between the target and the dump. The 
distance can however not be too small in order to ensure the necessary lateral 
separation between the  \PgSm-beam and the protons at the point where they are 
dumped. Taking this into account the distance was chosen to be 2.5 m. 

In the more downstream part of the vertical section "pockets" have been added. 
The purpose of these pockets is to diminish the number of interactions in the 
walls and to trap part of the showers in the tungsten.

There is a disadvantage with the wobbling configuration of the  beam as can be 
seen from the shape of the channel in the bend plane. Neutral particles can reach 
the exit of the channel from points far upstream in the channel.  The geometry 
actually allows neutrals to escape from the channel from interactions in the walls 
starting at a distance of only 3.5 m from the production target. A local photon 
background was observed in the experiment at a place corresponding to an angle 
of 7 mrad above the beam. As can be seen from Fig. \ref{fig:1c} this
is consistent with the geometry of the channel. To
 confirm that this neutral background originates from 
the upstream part of the channel, 2 radiation lengths of lead were introduced between 
the second and the third magnets. This reduced the background by a factor 2-3. 
Moreover, if the production target is removed, the background disappears 
completely, showing that the background does not originate from the
dump.

After the first beam period  in 1998-90 the shape of the channel was partially 
modified to diminish  this neutral background. The modifications are indicated in 
fig. \ref{fig:1c}. The second pocket was moved upstream and the walls were shaped in such 
a way that neutral particles from the target can not reach the
upstream wall of the pocket  which 
is seen by the exit and that the downstream part where neutral
particles may hit
the channel wall can not be seen from the exit. 
The point where the protons are dumped was also 
moved slightly  upstream in order to preserve the distance between the dump 
and  the second pocket to avoid any leakage into the pocket. The separation of the 
proton  beam and the \PgSm-beam at the dump was in this way reduced
from 8 mm to 6 mm.

Fig. \ref{fig:1e} shows a transversal  cross section of the
bending magnets. The pole gaps 
were filled with brass and tungsten collimators as mentioned
previously. Each magnet had a fixed part and a removable part of
inserts. The fixed part was made 
of brass for the two downstream magnets and of tungsten for the most upstream 
magnet. The removable part consisted of 90 tungsten block totally (30/magnet) 
with each block having its machined rectangular hole. The dimension of those 
blocks were 13 cm in the direction of the beam and $3.5 \times 18$ cm$^2$  transverse to the 
beam.

\subsection{Detection and identification of beam particles }\label{sec:detect}

The beam momentum was measured with three hodoscopes, namely one
scintillating fibre hodoscope placed before the last bending magnet
and two sets of $\mu$-strip silicon detectors located at two positions 
downstream of the magnetic channel (figure \ref{fig:1})
A scintillating fibre hodoscope was chosen to allow the   measurement
of  the particle position in the bending plane (z) with 
$ 250/\sqrt(12) \mu$m accuracy
in the area of strong stray magnetic field and in the strong radiation
environment of the magnetic channel.

The hodoscope was constructed of scintillating fibres 
(Kuraray SCSF-38) with  rectangular cross section (1 x 0.5 mm$^2$). 
The fibres were covered by a 55 $\mu$m thick cladding to which
a white pigment was  mixed to prevent crosstalk. 
Three fibres were put together 
to form one channel (3 $\times$ 0.5 mm$^2$, figure \ref{fig:1d}).
These three were glued to one clear fibre 
(1.5 mm $\times$ 1.5 mm $\times$ 1 m) as light guide  
which transmitted the light to one pixel of a 64-channel 
photo-multiplier. 
The channels were arranged in two rows which were displaced by half a 
fibre width (0.25 mm).
The hodoscope had a total of 64 channels, 32 in each row.
A valid hit was a coincidence between two fibres from each row. 
To prevent crosstalk at the tube causing wrong coincidences, neighbouring
fibres were distributed over the 64 channels of the PM.
Two such hodoscopes were used to obtain full efficiency.

At the exit of the beam channel the beam had to pass a hole of 
5 $\times$ 5 cm$^2$ defined by a veto counter consisting of 2 cm lead
sheets and 5 mm scintillator which was used to suppress the beam halo.

To reject  high momentum \Pgpm\ in the beam at an early  stage
of the trigger a fast transition radiation 
detector (TRD)   was placed approximately 1 m
downstream of the veto counter. This detector is described in detail
in \cite{BRU96}. Two scintillators each with an active area of 
$3.5 \times 3.5 $ cm $^2$ in coincidence placed in front  of the TRD and 
the experiments target were used to trigger on beam particles. 

The trajectory of the beam particle in front of the experimental target
was measured by a set of 8 high resolution silicon microstrip detectors 
with an active area of $ 5.1 \times 5.1 $ cm$^2$ and with a pitch of \mbox{
50 $\mu$m}. A set of 4 planes was placed in front of the TRD detector,
4 planes were placed behind the TRD directly  in front 
of the experiment target.
The detectors measured in two orthogonal directions. 
The lower part of Figure \ref{fig:1b} shows the setup around 
the experiment target. 

Particles emerging from the interaction were detected in the WA89
spectrometer shown in the upper part of figure \ref{fig:1b}. 
The secondary particles close to the interaction point were detected
by 20 silicon microstrip planes with 25 and 50$\mu $m pitch.
Two scintillators ($3.5 \times 3.5$ cm $^2$ each) 
were placed in the middle of the vertex detector
to measure the multiplicity of particles coming from the interaction. 
These detectors were used in the trigger. 

Positioning the target about 14~m
upstream of the centre of the $\Omega $-spectrometer provided a 10~m long
decay area for short-living strange particles. 
The products
of these decays along with the particles coming directly from the target
were detected by 40 planes of drift chambers with a spatial resolution
of about 300~$\mu $m. Special MWPC chambers (20 planes with 1mm wire spacing)  
were used in the central region of high particle fluxes.
In order to improve the track matching between the target region and the
decay region three sets of 4 MWPCs each with a pitch of 1 mm
were installed about 2 m behind the target.

The particle momenta were measured by the
$\Omega$-spectrometer  consisting of a super-con\-duc\-ting 
magnet with a field integral of 7.5~Tm and a tracking detector
consisting of 45 MWPC planes inside the field area and 12 drift chamber
planes at the exit of the magnet. The momentum resolution was
$\sigma (p)/p^{2}\approx 10^{-4}~(\gevc1)^{-1}$.

Charged particles were identified
using a ring imaging Cherenkov (RICH) detector . It had a 
threshold of $\gamma =42$  and provided $\pi/p$ separation
up to about 150~\gevc1. Downstream of the RICH a lead glass electromagnetic  
calorimeter was positioned for photon and electron detection .
This calorimeter was followed by a hadron calorimeter. For a more
detailed description see \cite{xipaper} and references therein. 

\section{Beam distributions }\label{sec:beamdis}

Figures \ref{fig:5}.c and \ref{fig:5}.d show  the beam profile in 
horizontal (y) and vertical (z) direction at the position of the experiment 
target about 3 m  from the exit of the beam channel. 
The beam had a width of 1.6 cm in y and a width of 3.7 cm in z
direction. The shape of the beam profile in z reflects the momentum
distribution of the beam. The beam profile in the non bending plane is
a result of the  profile of the proton beam and of the position of the
beryllium target. 

Figure \ref{fig:5}a and b show the correlations
between the beam particle impact point and its angle in the bending and
non bending plane. All particle 
trajectories in the narrow bands in figure \ref{fig:5} 
point back to the production target. Particles outside the correlation
bands are \Pgpm\ from \PgSm\ decays, particles elastically scattered
in the beam channel or muons. The distribution in figures \ref{fig:5}a
and \ref{fig:5}.b were used to measure the amount of  background in the
beam from these sources (see below). 

The momentum distribution of the beam was measured 
with the $\Omega$ spectrometer. 
For this measurement it was required that exactly 
one track was detected in the spectrometer to reject events with 
interactions. The momentum measurement for these non-interacting events 
was subsequently used to calibrate the beam hodoscope to be able to
measure the beam particle momenta also for events where the \PgSm\ 
interacted in the target. 
The momentum distribution  of beam particles for the 1994 data taking
period is shown 
in figure \ref{fig:6}. The solid line corresponds to the value measured
by the beam hodoscope. The dashed line shows the momentum measurement
in the spectrometer. This spectrum shows an additional component below
150 \gevc1\ stemming from the decays \PgSm\dcy\Pn\Pgpm.  
The measurement of the beam momentum gave average momenta of 345 \gevc1\
for the 1993 beam time and 330 \gevc1\ for the beam     time in 1994
and  a width of 
$\Delta$p/p = 18\% (FWHM) which  is close to the 
beam line design value of $\Delta$p/p = 15\%.

\section{Beam composition}\label{sec:beamcomp}

\subsection{\PgSm\ identification}\label{sec:sigma}

In the TRD high momentum \Pgpm\ were identified by their characteristic
transition radiation, which was detected in ten double planes of 
multiwire proportional chambers. If three of the ten planes 
gave a signal above threshold the beam particle was identified as a
high momentum pion and the event was subsequently rejected in the experiment trigger. 
The pulse height of each
chamber was recorded for monitoring
purposes and to allow a more detailed offline analysis at a later stage. 
Offline the pulse height information from the TRD chambers along  the 
track measured in  the the microstrip detectors 
was used to calculate a likelihood 
for the different particle hypotheses. The distribution of 
likelihoods for beam particles before and after
the TRD decision is shown in fig. \ref{fig:7}. 
From a two-component fit to the distribution of 
likelihoods before the TRD decision (fig \ref{fig:7}.a) a \PgSm/\Pgpm\ ratio of 
$ 0.38 \pm 0.04 $ at the position of the TRD can be deduced, which 
is in very good agreement with the design value of the beam line.
With the help of the TRD information the pion component for events in
the experiment trigger was reduced to a  \PgSm/\Pgpm\ ratio of $ 10
\pm 1  $ with a \PgSm\ detection efficiency of $ 82 \pm 1 \% $ 
By a cut on the TRD likelihood in the offline analysis a \PgSm\ purity
of 98.5 \% was achieved  while more than 95 \% of the \PgSm\ 
identified online were retained.

\subsection{\PgXm\ in the beam }\label{sec:cascades}

To measure the amount of \PgXm\ in the beam,  a sample
of events with beam particles passing through the target without
interaction was analysed. In this data sample we identified 
\PgSm\ and \PgXm\ decays by the
observation of decay  kinks  between the incoming 
beam particle and a negative particle detected in the spectrometer 
from the decays $ \Sigma^- \rightarrow n \pi^- $ and
$ \Xi^- \rightarrow \Lambda \pi^- $, respectively.
The decay lengths of \PgSm\ and \PgXm\ at the same momentum differ by
only $1\%$  and therefore  no acceptance or efficiency corrections 
are needed for a given fiducial volume  to evaluate the 
\mbox{\PgXm\ / \PgSm\ } ratio. 

In part a) of fig.~\ref{fig:8} the expected correlation (generated 
with the GEANT Monte Carlo) between momentum and kink angle 
 for \PgSm\ and \PgXm\ decays is shown.
The experimental data for the correlation are displayed in fig.~\ref{fig:8}b.
A second band originating from \PgXm\ decays can be discerned there.
For a quantitative analysis we projected this distribution  
along the angle-momentum correlation marked by the dashed and 
solid lines. This projection is shown by the solid 
histogram in part c) of fig.~\ref{fig:8}. 
Besides the dominating peak from \PgSm\ decays, a well
separated maximum corresponding to \PgXm\ decays can be identified.
The observed distribution can be reproduced by  
Monte Carlo simulations assuming a \PgXm\ contribution of 
(1.3 $\pm$ 0.1)\% relative to the number of \PgSm\  
(see cross-hatched histogram in fig.~\ref{fig:8}.c). This ratio is in
good agreement with the numbers given in \cite{cardello}.

\subsection{\PgOm\ and \PKm\ in the beam}\label{sec:OmK}

The production of \PgOm\ 
is expected to be even more suppressed than the production of \PgXm\
due to the additional unit of strangeness. In figure
\ref{fig:8}b  a sizeable contribution of \PgOm\ would be visible 
in a band above the correlation for \PgSm\ in the momentum range of
70-160 \gevc1 . In this region of the plot no \PgOm\ can be identified
above the background. The expected \PgOm/\PgXm\ ratio at the channel
exit calculated from the numbers given in \cite{cardello} is 
$ 2 \times 10^{-3} $ which is below the sensitivity of the 
measurement presented in figure  \ref{fig:8}.

The \PKm\ component of the beam can be estimated on the basis of the
\PKm\ to \Pgpm\ production ratio, which was measured to be $(1.15 \pm
0.02)\% $ \cite{ATH80} in 300 
\gevc1\ p-Be collisions . Since the \PKm\ are below the TRD threshold
they were not rejected by this detector. From the \PKm/\Pgpm\ ratio at
production we deduce that there was a 
$\rm 2.1\%$ admixture of \PKm\ in the beam after the 
online rejection of the high-momentum \Pgpm\ component. 

\subsection{Muon halo}\label{sec:muon}

There are two different origins of the muon background. One part comes from 
showers created by the protons being dumped in the tungsten. The second part is 
due to secondary pions and kaons created in the production target which decay 
into muons before being absorbed in the walls of the channel.
The first kind of muon background is minimized by choosing a very dense 
material  with a very short interaction length, such as tungsten. The second 
contribution depends on the geometry of the beginning of the channel as 
discussed in section \ref{sec:magnetic}. To obtain a rough idea of the
muon background we have 
used the Monte-Carlo program HALO \cite{HALO}. This program generates parent 
particles decaying into muons and the muons are then tracked through the  coils 
and  yokes of the different magnets. The result of  this calculation is shown in 
table \ref{tab:2}, where muon densities  for $2\times 10^{10}$
protons  are given at 
the entrance to the Omega spectrometer. Contributions from the dump and from 
the target have been separated. It can be seen that at the beam axis we expect 
about $2 \times  10^5$ muons/m$^2$ and that the dominating part comes from
secondaries  from the target.

The level of the muon background was measured with scintillator hodoscopes 
located a couple of meters downstream of the exit of the Omega spectrometer. 
About $2\times 10^5$ muons/m$^2$ per pulse were measured, 
in good agreement with the 
calculations. Taking out the production target reduced this rate by a 
factor of 2.5, confirming the trends found in the calculations.

\subsection{Backgrounds in the beam }

The total amount of background in the beam can be estimated
from the plots in figure \ref{fig:5}. In this plots about 21 \% of all
measured particle trajectories do not fall into the angle-impact
correlation which is given for particles coming from the production
target. Of this 21 \% about half could be assigned to \Pgpm\ from \PgSm\
decays as shown by  a Monte Carlo simulation of the beam line.  
The rest corresponds to muons and particles elastically scattered 
in the collimators. 

To measure the effect of the veto counter data were taken in a special
run where the veto counter was not included in the experiment
trigger. Removing the veto counter resulted in an increase of 5 \% of 
the  multiplicity dependent part of the trigger rate. This
increase was caused by additional particles crossing the apparatus 
together with the particles produced in the experiment target  causing
larger pulse heights in the multiplicity counters.

\subsection{ Interactions from Pions and Neutrons }

With the given beam composition it was possible to obtain 
in addition to the events with \PgSm\ induced interactions 
data samples on interactions induced by \Pgpm\ and neutrons: 

To obtain a clean sample of events with pion interaction 
data was taken with an inverted  TRD decision for a small 
fraction of the runing time.
In this data sample the \PgSm\-component
of the beam was suppressed online by the TRD to a 
level below 2 \%. 

For \PgSm\ decaying upstream of the centre of the  last bending 
magnet the decay pion fell in the acceptance of the 
beam trigger counters. Since the 
momentum of the decay pion is below 150 \gevc1\ this events were not
rejected by the TRD  and were identified with the
angle-impact correlation as mentioned above. In some fraction of this
events however an interaction was induced by the decay neutron. 
It was possible to isolate a clean sample of 
such events  by requiring that the beam particle trajectory was
outside the angle-impact correlation and not pointing to the
reconstructed interaction vertex. Additionaly it was required that the
pion track could be followed  into the spectrometer and had a momentum 
below 150 \gevc1.

Neutron and \Pgpm\ induced interactions were used as control
data sets to be able to compare to other experiments using these 
beams types \cite{xipaper}. 

\section{ Effect of the vertical target position }     

The vertical target position determined the position of 
the produced particles at the entrance of the magnetic channel and
therefore the curvature of the central trajectory passing
through the collimators of the magnetic channel. By 
moving the target up and down it was hence possible to tune  
the average beam momentum.  For each given target
position the  proton beam could be steered onto the production 
target by means of a trimmer magnet. 

The target position with  maximum channel transmission 
was given by the geometrical acceptance of the collimator system. 
Since the \PgSm/\Pgpm\ production ratio is
strongly dependent on the average beam momentum this 
target steering also had an impact on the \PgSm\ content of the beam. 

To find an optimal point of operation the target was moved 
$\pm$ 0.4 cm relative to its nominal zero position and the
corresponding  beam momenta and particle fluxes were measured. 
Figure \ref{fig:9} shows the results of these measurements.
The  average particle momentum varied from 310 to 360 \gevc1\ for 
this range of target positions (fig \ref{fig:9}.a). The flux variation
is shown in figure \ref{fig:9}.b .  At maximum transmission the
flux was  $5 \times 10^5$ particles (\PgSm\ and \Pgpm\ )
 for $3.0 \times 10^{10}$ incident protons.
In 1993 the beam was operated with the target at position -0.05  which gave an 
average momentum of 345 \gevc1, a \PgSm/\Pgpm\ ratio of 
0.4 and a flux of $1.38 \times 10^5$ \PgSm\ particles at the channel exit.

\section{Conclusions }

A high-intensity double-bend hyperon beam was successfully
operated at momenta between 330 and 345 \gevc1\ 
delivering $ 1.4 \times 10^5$ \PgSm\ to 
experiment WA89  in the CERN West Area. At the experiment target
3 m downstream of the channel exit the beam had a \PgSm/\Pgpm\ ratio of
$0.38 \pm 0.04 $ and a size of 1.6 $\times $ 3.7 cm$^2$.  The 
\PgXm/\PgSm\ component of the beam was measured to be $ 1.3 \pm 0.1  $
\% in good agreement with previous measurements. The muon and 
neutral backgrounds were substantially reduced by a careful design of
the proton beam dump area and the shape of the collimators. 
By an adjustment of the vertical target position it was possible to 
tune the beam momentum, the particle flux and the \PgSm/\Pgpm ratio 
to an optimal value. 

\section{Acknowledgement}

The members of the WA89 collaboration are grateful to the EA group
for their continuous support in the construction and operation of the
hyperon beam line, to the SPS staff 
for providing good beam conditions and to the Omega group 
for their help in running the Omega spectrometer. We would like to
thank J. Zimmer for his invaluable support during all phases of
detector development and setup.

%

\begin{table}[htbp]
  \begin{center}
    \leavevmode
    \begin{tabular}[h]{|ll|l|}\hline
      Length of the beam from target && \\ 
      to exit of channel (m)         && 12.7 \\ \hline
      B $\times$ l (Tm)              && +17.2 ; -8.6 \\ \hline
      Deflection (mrad)              && +13.8 ; -6.9 \\ \hline
      Central  momentum (GeV/c)          && 355 \\ \hline
      Acceptance ($\pm2 \sigma$)       &$\alpha_H$ (mrad)& $\pm 1.5$ \\
                                     &$\alpha_V$ (mrad)& $\pm 0.6$ \\
                                     &$\alpha_V$ (mrad)& $\pm 12 $ \\ \hline
   Spot size at exit ($x \times y $\ mm$^2$)&& $37 \times 16 $\\ \hline
   Flux of \PgSm\ for $3 \times 10^{10}$ ppp && $1.4 \times 10^5$ \\ \hline
   \PgSm/\Pgpm\ ratio at channel exit  && 0.38 $\pm$  0.04 \\ \hline
    \end{tabular}
    \caption{Parameters of the \PgSm\  beam}
    \label{tab:1}
  \end{center}
\end{table}

\begin{table}[htbp]
  \begin{center}
    \leavevmode
      \begin{tabular}[h]{|l|c|c|c|c|} \hline
   & \multicolumn{2}{|l|}{ Muons/m$^2$  at the beam axis}    & \multicolumn{2}{|l|}{Muons/m$^2$ 0.85 m above (\Pgmm)}\\
   & \multicolumn{2}{|l|}{ over an area of $1 \times 1 $m$^2$} & \multicolumn{2}{|l|}{and below (\Pgmp) the beam axis} \\
   & \multicolumn{2}{|l|}{                                 } & \multicolumn{2}{|l|}{over an area of 0.3 $\times$ 0.3 m$^2$} \\ \hline
   & \Pgmp         & \Pgmm    & \Pgmp         & \Pgmm \\ \hline
From target &$ 2.2\times 10^4 $&$ 1.3\times 10^5$&$1.3\times 10^6$&$ 9.6\times 10^5$ \\
From dump   &$ 0.3\times 10^4 $&$ 0.2\times 10^5$&$ 0.1\times 10^6$&$ 0.5\times 10^5$ \\
Total       &$ 2.5\times 10^4 $&$ 1.5\times 10^5$&$ 1.4\times 10^6$&$ 1.0\times 10^6 $\\ \hline
      \end{tabular}
    \caption{Calculated muons/m$^2$ at the entrance of the Omega
      spectrometer for $2 \times 10^{10}$ incident protons}
    \label{tab:2}
  \end{center}
\end{table}

%
%
\begin{figure}[H]
\begin{center}
\mbox{\epsfxsize=13cm\epsffile{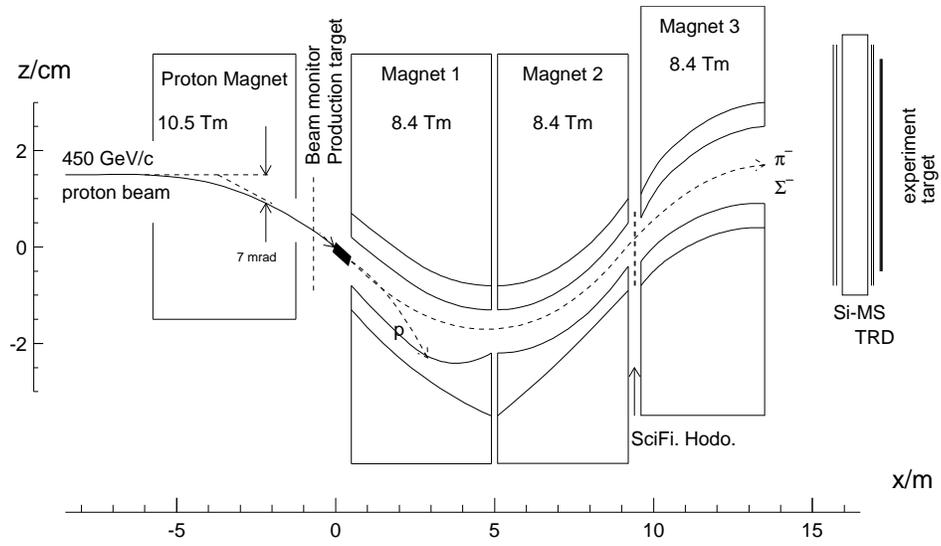}}
\vspace{0cm}
\caption{Basic layout of the hyperon beam channel and the target area 
of  experiment WA89}
\label{fig:1}
\end{center}
\end{figure}

%
%
\begin{figure}[H]
\begin{center}
\mbox{\epsfxsize=11cm\epsffile{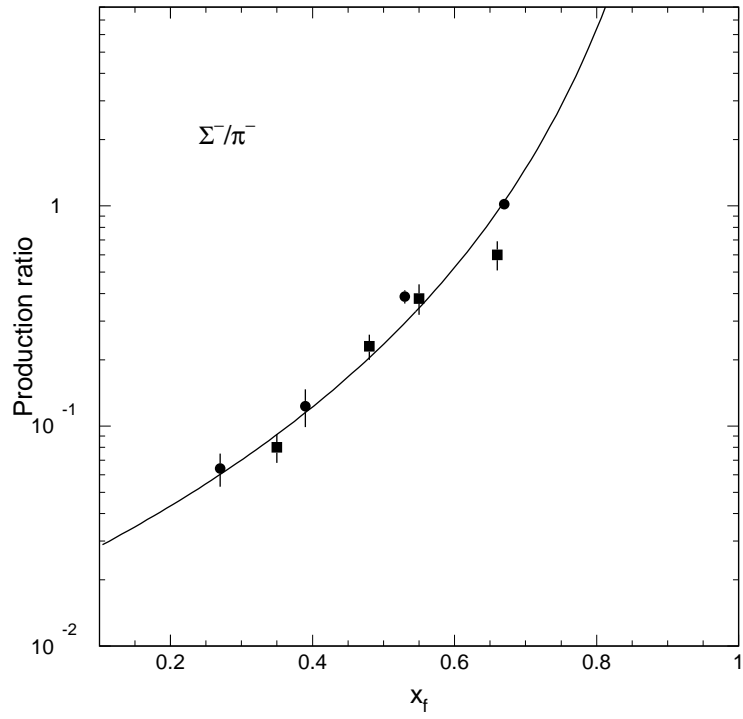}}
\caption{The \PgSm/\Pgpm production ratio at small $p_t$ 
 as function of $ \xf$  (Solid squares from ref. \cite{Bour}; 
Solid circles from ref. \cite{cardello}). The solid line is a 
fit to the data at $p_t=0$ given in \cite{cardello}.}
\label{fig:2}
\end{center}
\end{figure}

%
%
\begin{figure}[H]
\begin{center}
\mbox{\epsfxsize=13cm\epsffile{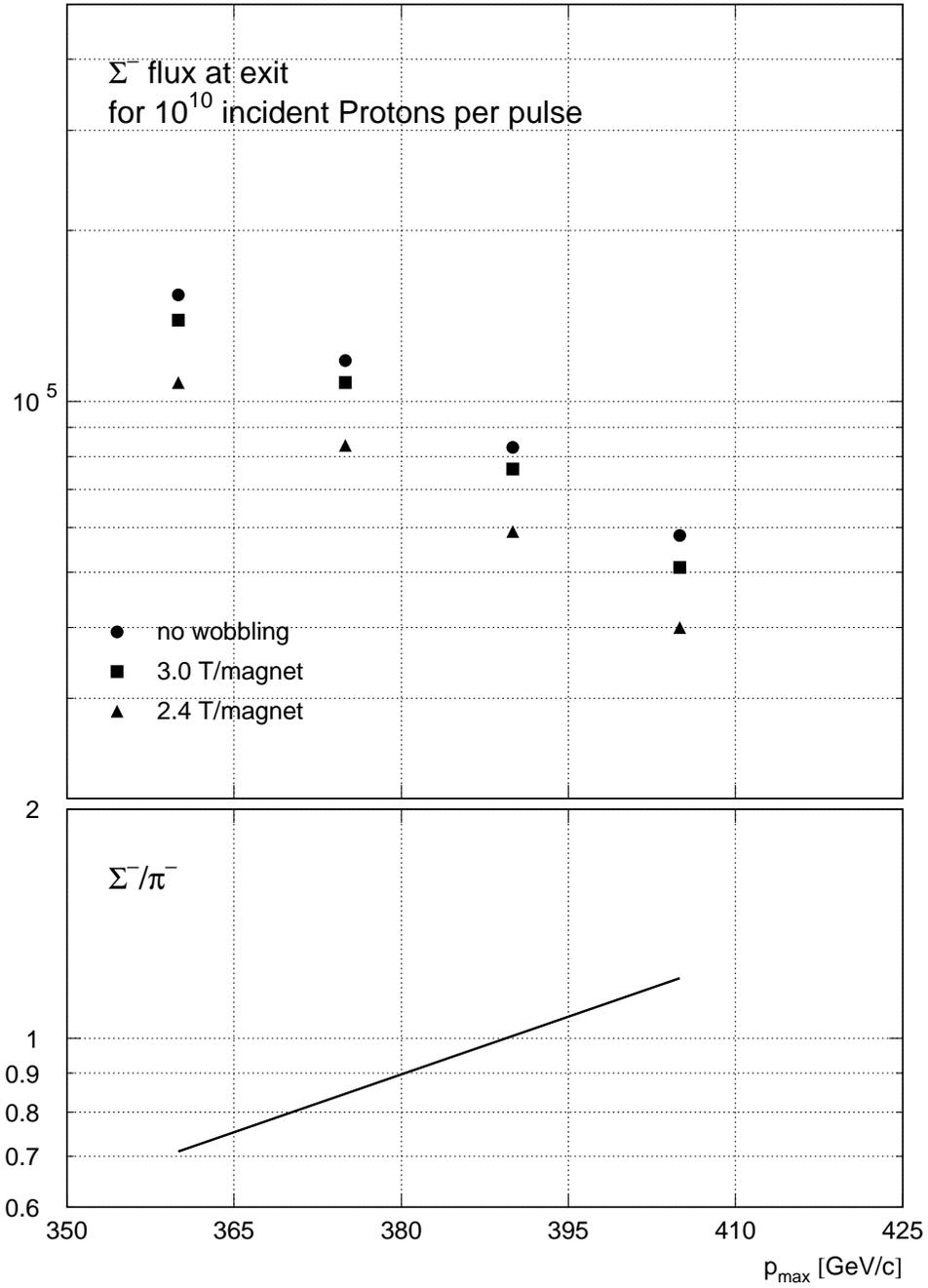}}
\vspace{0cm}
\caption{\PgSm-flux for $2 \times 10^{10}$ protons per pulse as a
  function of momentum of the central beam trajectory. 
  All acceptances were adjusted to give $\Delta p/ p = 15 \%$ (FWHM)}
\label{fig:3}
\end{center}
\end{figure}

%
%
\begin{figure}[H]
\begin{center}
\mbox{\epsfxsize=13cm\epsffile{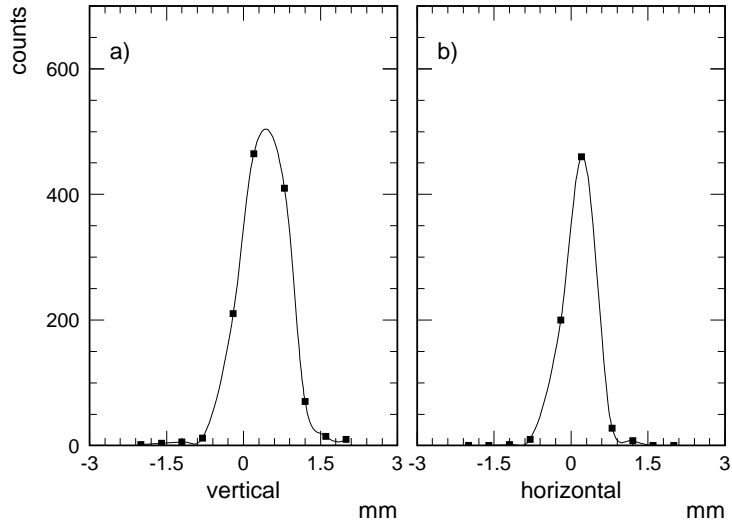}}
\vspace{0cm}
\caption{Vertical (a) and horizontal (b) beam profile of the proton
  beam at the production target }
\label{fig:4}
\end{center}
\end{figure}

\vspace{-3cm}

%
%
\begin{figure}[H]
\begin{center}
\mbox{\epsfxsize=12cm\epsffile{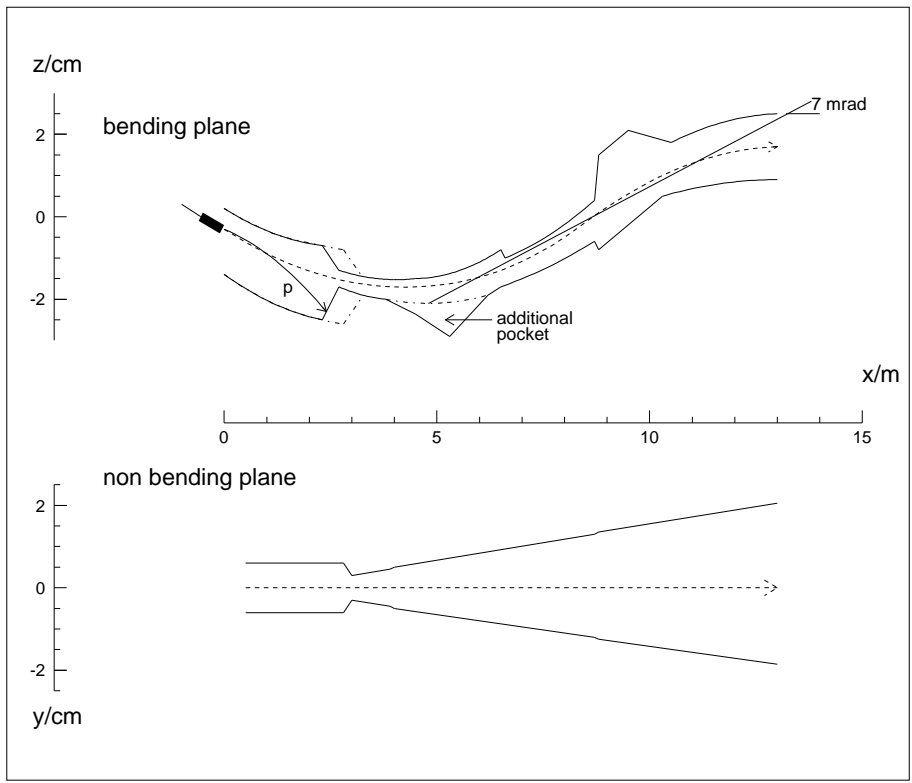}}
\caption{Shape of the tungsten channel in the horizontal and vertical
  plane. The dashed contours correspond to the pre-1991 setup}
\label{fig:1c}
\end{center}
\end{figure}

%
%
\begin{figure}[H]
\begin{center}
\mbox{\epsfxsize=10cm\epsffile{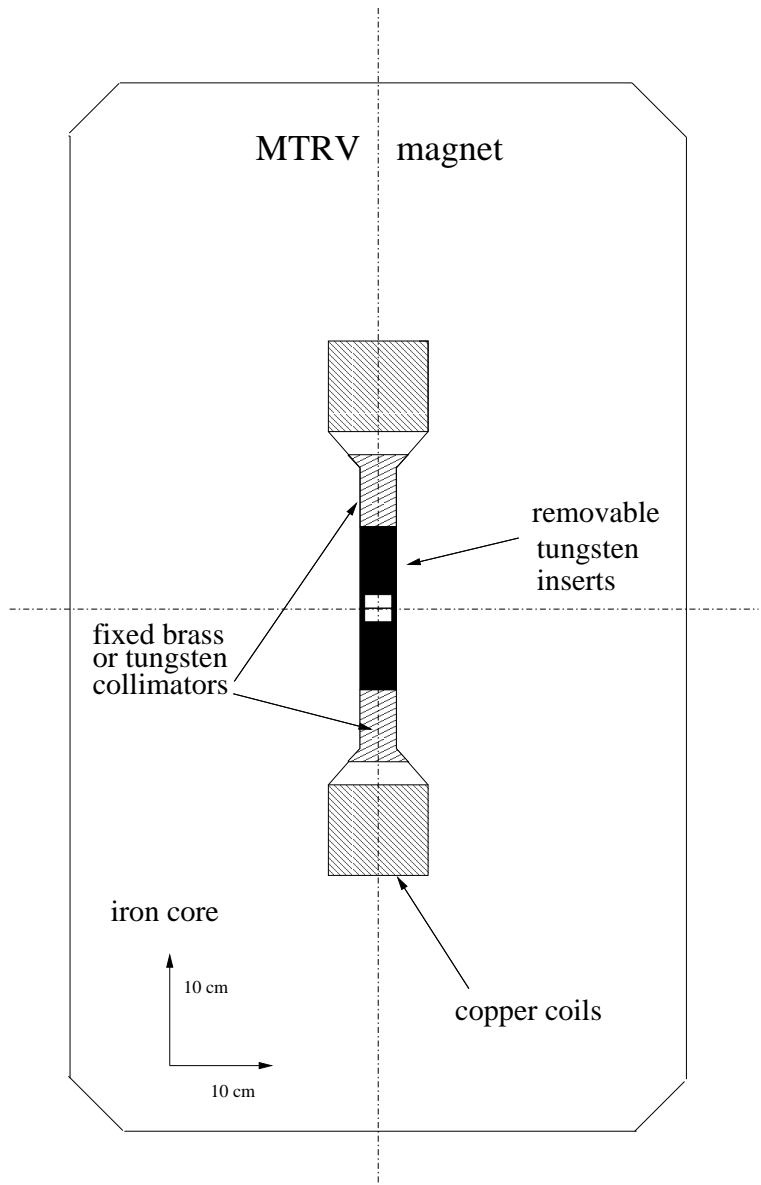}}
\caption{Transverse cross section of a bending magnet}
\label{fig:1e}
\end{center}
\end{figure}

%
%
 \begin{figure}[H]
\begin{center}
\mbox{\epsfxsize=11.cm \epsffile{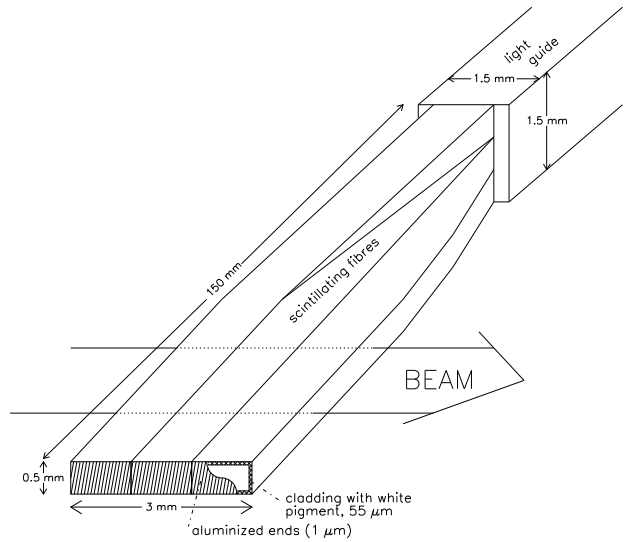}}

\caption { One channel of the beam hodoscope. The scintillating fibres were 
glued to a light guide (1 m long) which leads the photons to the PMT. 
The open ends of the scintillating fibres were aluminized  to reflect the 
light emitted in the direction opposite to the light guide.}
\label{fig:1d}
\end{center}
\end{figure}

%
%
\begin{figure}[H]
\mbox{\epsfxsize=11cm\epsffile{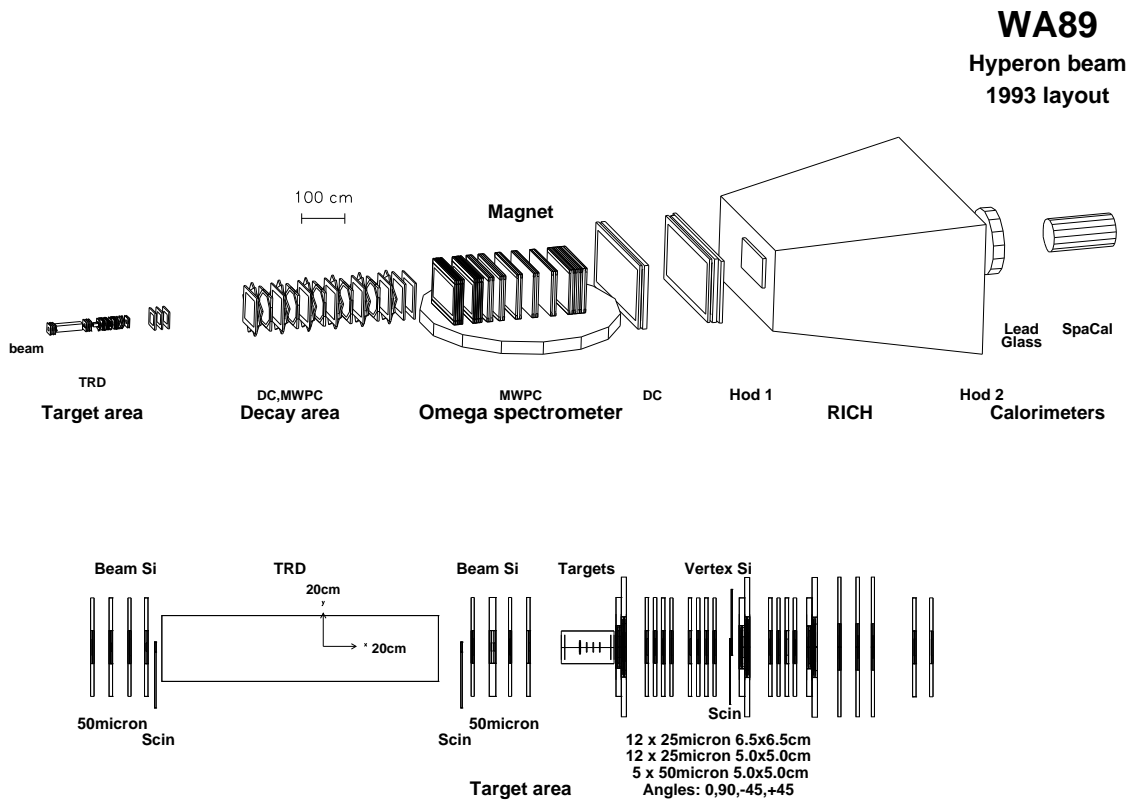}}
\caption{Setup of experiment  WA89. The lower part shows the
  expanded view of the target area}
\label{fig:1b}
\end{figure}

%
%
\begin{figure}[H]
\begin{center}
\mbox{\epsfxsize=13cm\epsffile{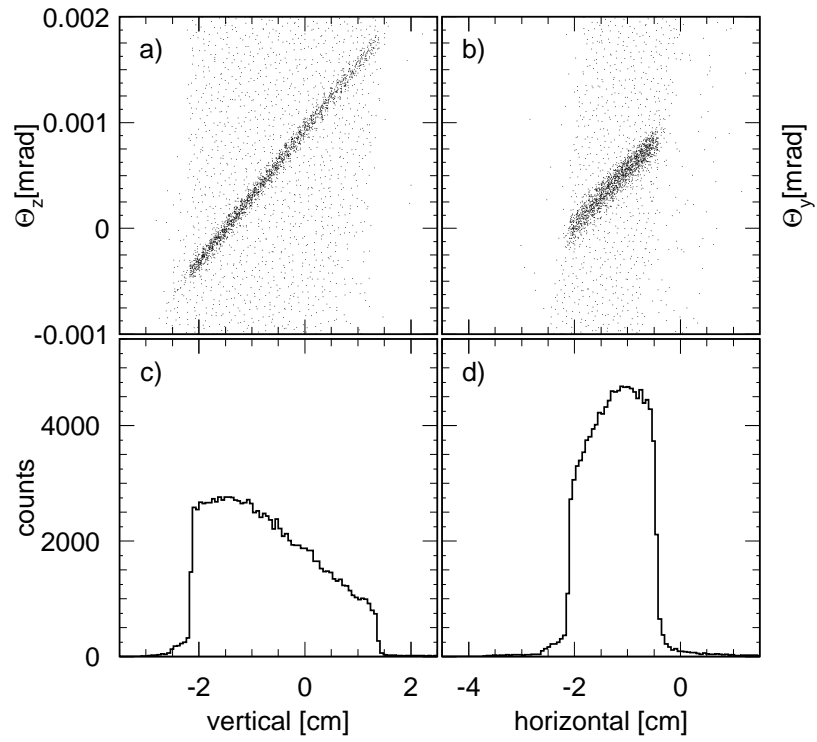}}
\vspace{0cm}
\caption{Angle-impact correlations ( a and b ) and beam profiles (
  c and d) at the experiment target $\sim$3 m downstream from the 
channel exit. }
\label{fig:5}
\end{center}
\end{figure}

%
\begin{figure}[H]
\begin{center}
\mbox{\epsfxsize=13cm\epsffile{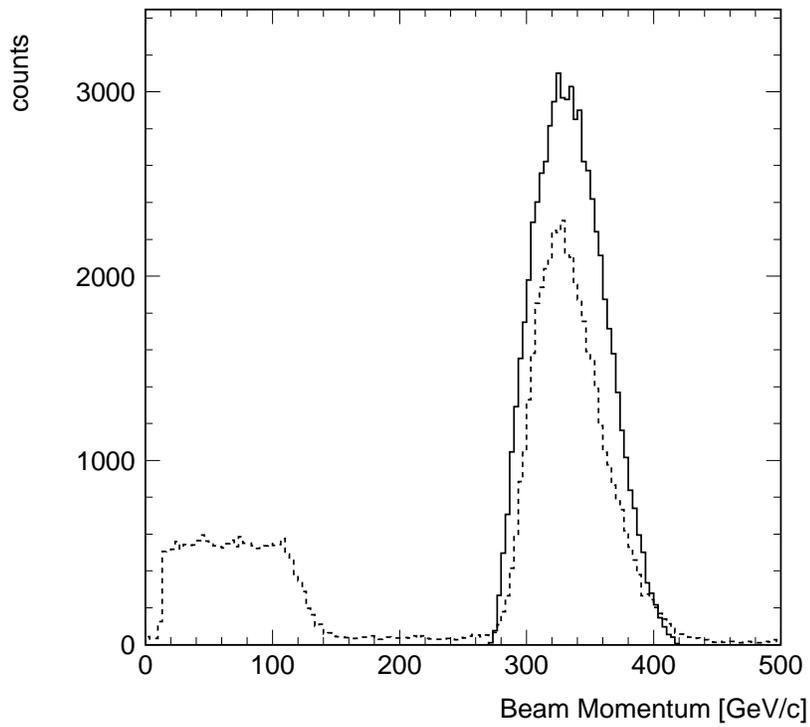}}
\caption{Beam momentum spectrum as measured with the
  beam hodoscope (solid line) and as measured in the 
 $\Omega$-Spectrometer (dashed line). The additional component below
 150 \gevc1\ stems from decays \PgSm\dcy\Pn\Pgpm (Data from 1994 beam time).}
\label{fig:6}
\end{center}
\end{figure}

%
%
\begin{figure}[H]
\begin{center}
\mbox{\epsfxsize=10cm\epsffile{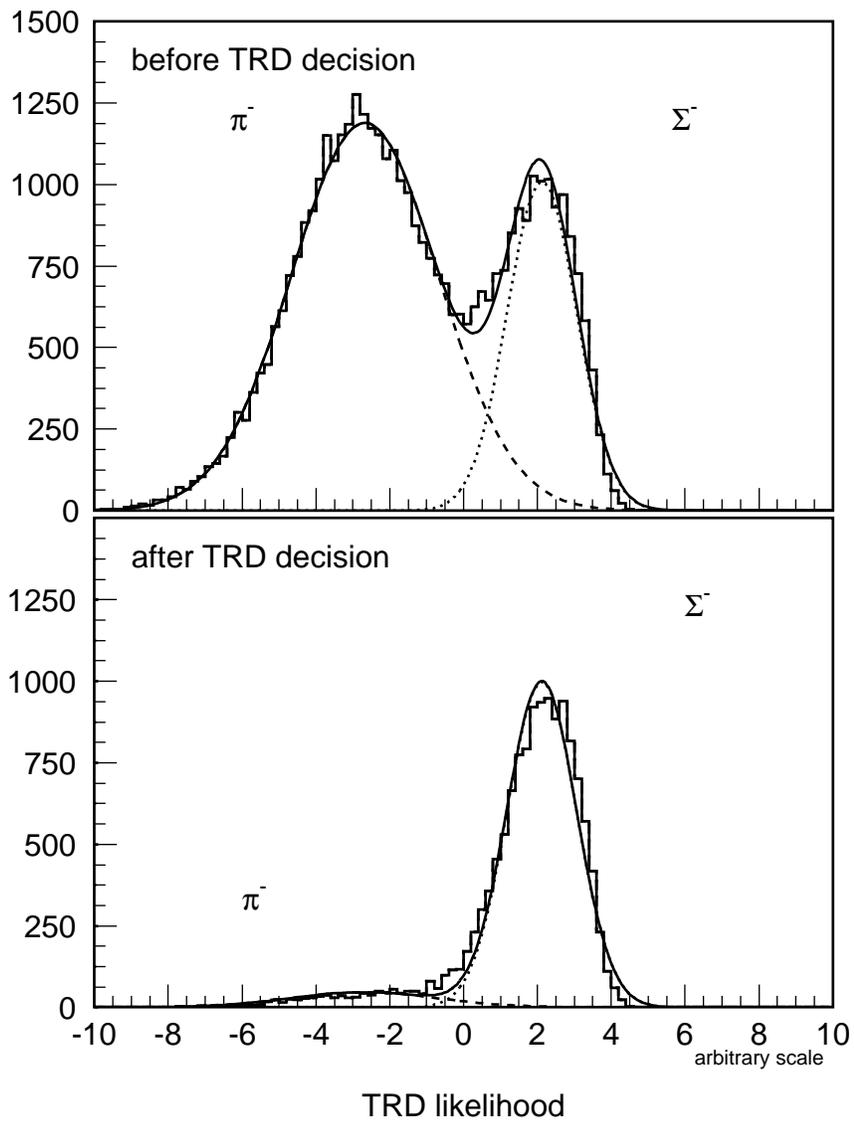}}
\caption{Distributions of TRD likelihoods before and 
after the TRD decision}
\label{fig:7}
\end{center}
\end{figure}

%
%
\begin{figure}[H]
\begin{center}
\mbox{\epsfxsize=13cm\epsffile{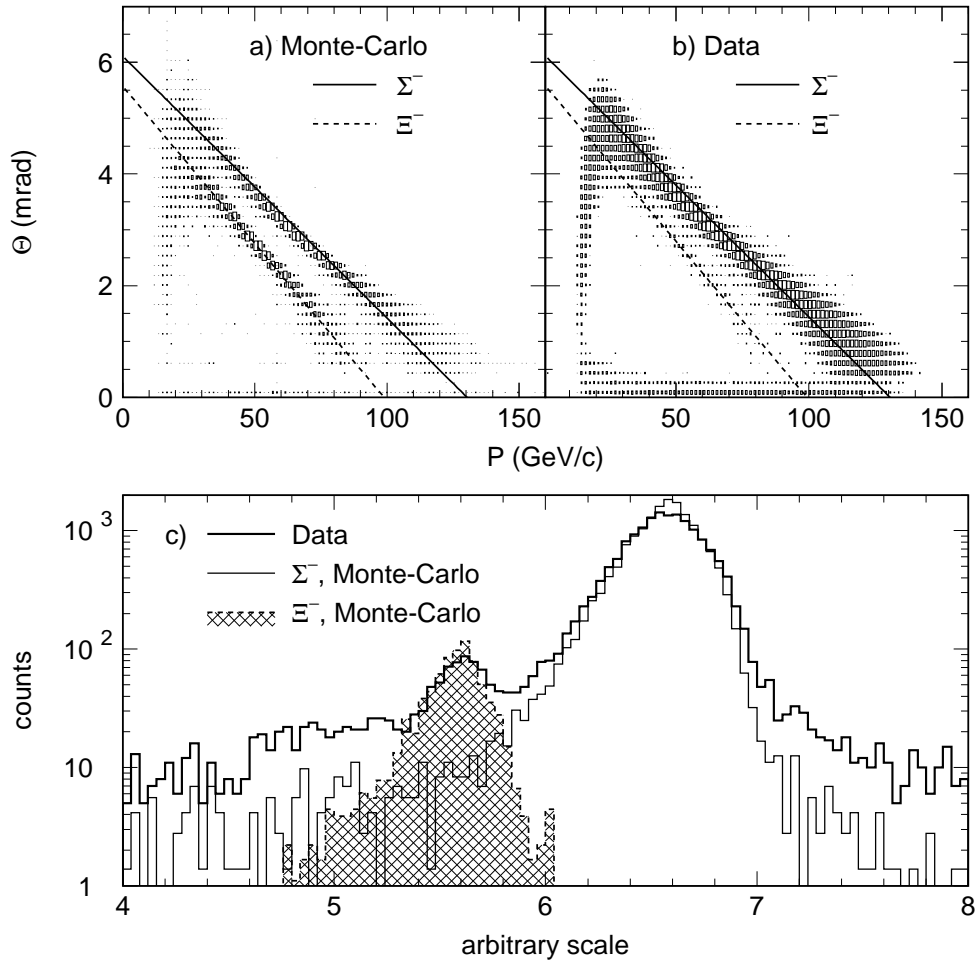}}
\vspace{0cm}
\caption{Correlation plots between momentum and decay angle between 
incoming \PgSm\ and \PgXm\ and outgoing \Pgpm\ for a) Monte
Carlo and b) Data. The projections of these distributions along the 
two indicated lines are shown in part c).\cite{xipaper}}
\label{fig:8}
\end{center}
\end{figure}

%
%
\begin{figure}[H]
\begin{center}
\mbox{\epsfxsize=12cm\epsffile{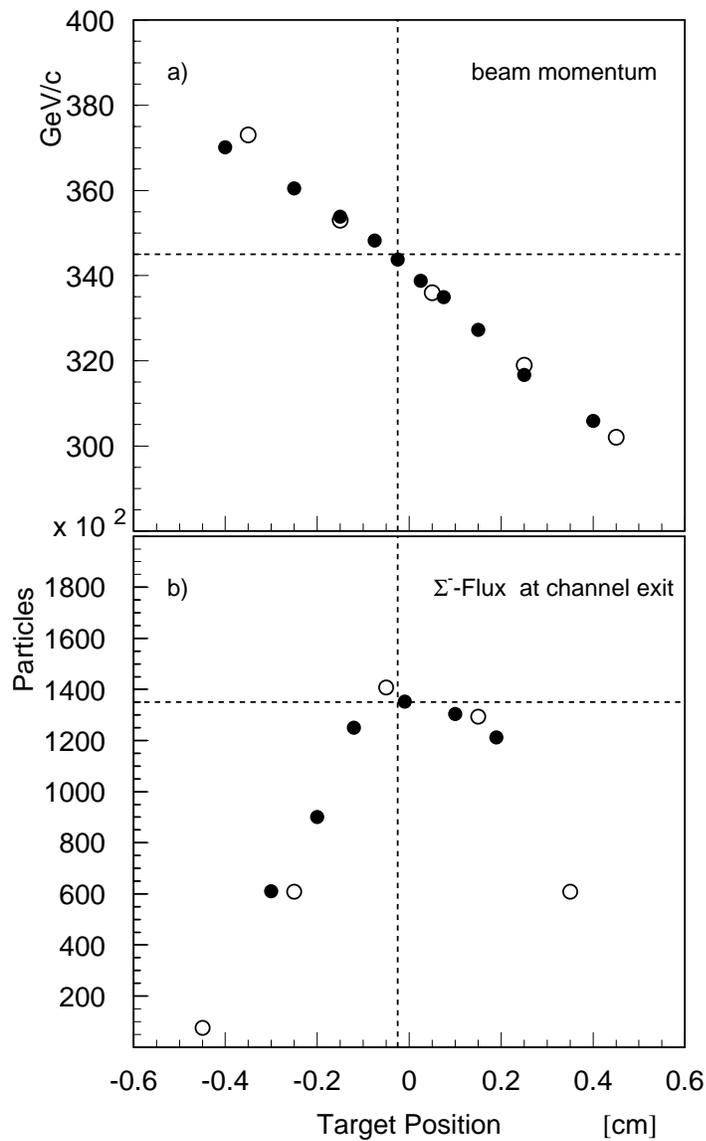}}
\vspace{0cm}
\caption{Beam momentum  and  particle flux as function 
of the target position. The solid points correspond to the
measurements, 
the open points to the MC-calculation. The dashed lines indicate the 
working point of the hyperon beam channel in 1993.  }
\label{fig:9}
\end{center}
\end{figure}

\end{document}